{}%\documentclass[aps,prb,preprint,showpacs,preprintnumbers,superscriptaddress,linenumbers]{revtex4}
\documentclass[aps,prb,reprint,superscriptaddress]{revtex4-1}
\usepackage{graphicx}
\usepackage{bm}
\usepackage{color}

\usepackage{units,amsmath,amsfonts,amssymb}
\begin{document}
\title{First-order multi-k phase transitions and magnetoelectric effects in multiferroic Co$_3$TeO$_6$}

\author{Pierre Tol\'{e}dano}
 \affiliation{Laboratory of Physics of Complex Systems, University of Picardie, 33 rue Saint-Leu, 80000 Amiens, France}
 \affiliation{HISKP, Universit\"at Bonn, Nussallee 14-16, D-53115 Bonn, Germany}
\author{Vera Carolus}
 \affiliation{HISKP, Universit\"at Bonn, Nussallee 14-16, D-53115 Bonn, Germany}
\author{Matthias Hudl}
 \affiliation{Department of Engineering Sciences, Uppsala University, Box 534, SE-751 21 Uppsala, Sweden}
\author{Thomas Lottermoser}
 \affiliation{Department of Materials, ETH Zurich, Wolfgang-Pauli-Strasse 10, 8093 Zurich, Switzerland}
\author{Dmitry D. Khalyavin}
 \affiliation{ISIS facility, STFC Rutherford Appleton Laboratory, Chilton, Didcot, Oxfordshire, OX11-0QX, United Kingdom}
\author{Sergey A. Ivanov}
 \affiliation{Department of Engineering Sciences, Uppsala University, Box 534, SE-751 21 Uppsala, Sweden}
 \affiliation{Department of Inorganic Materials, Karpov Institute of Physical Chemistry, Vorontsovo pole, 10, 105064, Moscow K-64, Russia}
\author{Manfred Fiebig}
 \affiliation{Department of Materials, ETH Zurich, Wolfgang-Pauli-Strasse 10, 8093 Zurich, Switzerland}

\date{\today}

\begin{abstract}
A theoretical description of the sequence of magnetic phases in Co$_3$TeO$_6$ is presented. The strongly
first-order character of the transition to the commensurate multiferroic ground state, induced by
coupled order parameters corresponding to different wavevectors, is related to a large magnetoelastic
effect with an exchange energy critically sensitive to the interatomic spacing. The monoclinic magnetic
symmetry $\text{C}2'$ of the multiferroic phase permits spontaneous polarization and magnetization as
well as the linear magnetoelectric effect. The existence of weakly ferromagnetic domains is verified
experimentally by second harmonic generation measurements.
\end{abstract}

%\pacs{}
%\keywords{}

\maketitle

At variance with structural transitions which alter the lengths and orientations of the chemical
bonds, the spin ordering occurring in magnetic phases has in most cases a negligible effect on the
structural lattice. An important exception is represented by the class of magnetostructural
transitions occurring in multiferroic compounds in which the magnetic ordering in the multiferroic
phase induces simultaneously a change in the atomic structure, which permits the emergence of a
spontaneous polarization.\cite{1,2} However, the measured changes of lattice parameters found in
the multiferroic phases are generally small, i.e. of the order of $\unit[10^{-3}]{\text{\AA}}$,
and do not affect the second-order  character of the transitions to these phases.\cite{3,4} Here,
we describe theoretically the sequence of phases recently reported in Co$_3$TeO$_6$\cite{5,6} in
which a strongly first order transition, characterized by substantial discontinuities of the
lattice parameters and a remarkable delta-shape peak of the specific heat, yields a multiferroic
ground state displaying magnetoelectric properties. The observed structural changes are related to
the coupling between the magnetic order-parameters involved at the transition, which correspond to
different propagation wave-vectors, in contrast with the standard situation found in multiferroic
transitions where the coupled order-parameters generally pertain to the same k-vector.\cite{3,8n}

\begin{figure}[htb]
\includegraphics[width=0.75\columnwidth]{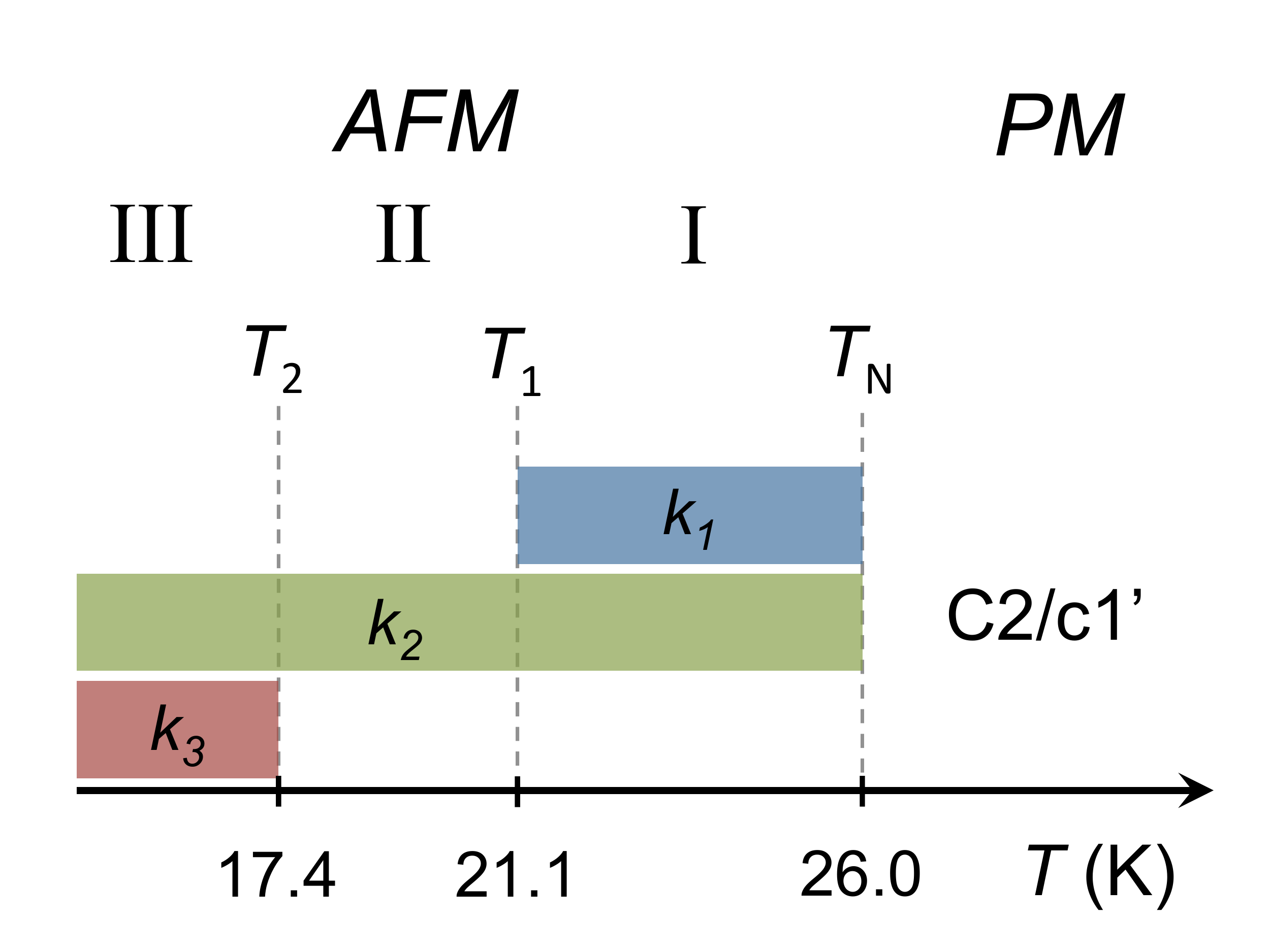}
\caption{\label{fig3} Sequence of phases observed in Co$_3$TeO$_6$ from Ref.~\onlinecite{6}}
\end{figure}

Neutron powder diffraction studies\cite{6,6n} show that below the paramagnetic phase described by
the space group $G_P = \text{C}2/\text{c}1'$ the Co$_3$TeO$_6$ undergoes a sequence of three
antiferromagnetic phases, summarized in Fig.~\ref{fig3}.\cite{footnote-CTO} They are associated
with three different k-vectors of the centred monoclinic Brillouin-zone: $\vec{k}_1 = (0, 0.480,
0.055)$, $\vec{k}_2 = (0, 0, 0)$ and $\vec{k}_3 = (0, 1/2, 1/4)$.
%\section{}

I. The incommensurate Phase I, which emerges at $T_N = \unit[26]{K}$, shows the coexistence of
$\vec{k}_1$ and $\vec{k}_2$. The transition order-parameter transforming as the irreducible
representation (IR) $\tau_1 (\vec{k}_1)$ (see Table~\ref{table1})\cite{9} has four components
$(\eta_1 = \rho_1 e^{i \theta_1}$, $\eta_1^* = \rho_1 e^{-i \theta_1}$, $\eta_2 = \rho_2 e^{i
\theta_2}$, $\eta_2^* = \rho_2 e^{-i \theta_1})$ which yield the free-energy
\begin{equation}
 F_1(\rho_1, \rho_2) =
 \frac{\alpha_1}{2}(\rho_1^2+\rho_2^2)+\frac{\beta_1}{4}(\rho_1^4+\rho_2^4)+\frac{\beta_2}{2}\rho_1^2\rho_2^2.
\end{equation}
Minimizing $F_1$ leads to two possibly stable states having the same point-group symmetry
$2/\text{m}1'$ as the paramagnetic phase, and corresponding to the equilibrium conditions $(\rho_1
\neq 0, \rho_2 = 0)$ and $(\rho_1 = \pm \rho_2)$ . The point group symmetry $2/\text{m}1'$ of
phase I and its incommensurate character are preserved when considering a coupling with the
order-parameter associated with $\vec{k}_2$.
\begin{table*}
\caption{\label{table1} Generators of the active irreducible representations of the paramagnetic
space-group $\text{C}2/\text{c}1'$ associated with the wave-vectors $\vec{k}_1$, $\vec{k}_2$ and
$\vec{k}_3$ in Co$_3$TeO$_6$. Columns matrices represent diagonal matrices. $T$ is the
time-reversal operator. $\epsilon = \exp{({\rm i}k_1^z c)}$, $\omega = \exp{({\rm i}k_1^y b/2)}$}
\begin{ruledtabular}
\begin{tabular}{cccccc}
& $(2_y \left|\right. 0,0,\frac{c}{2})$ & $(\bar{1} \left|\right. 0,0,0)$ & $T$ & $(1 \left|\right. \frac{a}{2},\frac{b}{2},0)$ & $(1 \left|\right. 0,0,c)$\\
$\tau_1(\vec{k}_1)$
&
$\begin{bmatrix} & & 1 & \\ & & & 1\\ 1 & & & \\ & 1 & &  \end{bmatrix}$ &
$\begin{bmatrix} & 1 & & \\ 1& & & \\ & & & \epsilon^*\\ & & \epsilon &  \end{bmatrix}$
&
$\begin{bmatrix} -1 \\ -1 \\ -1 \\ -1  \end{bmatrix}$
&
$\begin{bmatrix} \omega \\ \omega^* \\ \omega \\ \omega^*  \end{bmatrix}$
&
$\begin{bmatrix} \epsilon \\ \epsilon^* \\ \epsilon^* \\ \epsilon  \end{bmatrix}$
\\
$\Gamma_4(\vec{k}_2)$ & $\mathbf{-1}$ & $\mathbf{-1}$ & $\mathbf{-1}$ & $\mathbf{1}$ & $\mathbf{1}$\\
$\tau_1(\vec{k}_3)$&
$\begin{bmatrix} & & 1 & \\ & & & 1\\ 1 & & & \\ & 1 & &  \end{bmatrix}$ &
$\begin{bmatrix} & 1 & & \\ 1& & & \\ & & & -i \\ & & i &  \end{bmatrix}$
&
$\begin{bmatrix} -1 \\ -1 \\ -1 \\ -1  \end{bmatrix}$
&
$\begin{bmatrix} i \\ -i \\ i \\ -i  \end{bmatrix}$
&
$\begin{bmatrix} i \\ -i \\ -i \\ i  \end{bmatrix}$
\end{tabular}
\end{ruledtabular}
\end{table*}

II. The commensurate phase II appearing at $T_1 = \unit[21.1]{K}$ below a second-order transition
corresponds to the single wave-vector $\vec{k}_2$. It remains stable in a narrow interval of
temperature down to $T_2 = \unit[17.4]{K}$. The one-dimensional IRs $\Gamma_1-\Gamma_4$ at the
centre of the monoclinic Brillouin zone\cite{9} induce respectively the magnetic symmetries
$\text{C}2/\text{c}$ ($\Gamma_1$), $\text{C}2'/\text{c}'$ ($\Gamma_2$), $\text{C}2/\text{c}'$
($\Gamma_3$) and $\text{C}2'/\text{c}$ ($\Gamma_4$). The magnetic structure proposed by Ivanov et
al.\cite{6} from neutron data coincides with the $\text{C}2'/\text{c}$ magnetic group which is
therefore associated with a one-dimensional order-parameter, denoted $\zeta$ hereafter,
corresponding to the single equilibrium state which minimizes below $T_1$ the canonical
free-energy:
\begin{equation}
 F_2(\zeta) = \frac{\alpha_2}{2}\zeta^2 + \frac{\lambda_1}{4}\zeta^4
\end{equation}
In absence of applied fields the magnetic symmetry of the phase does not allow the emergence of
spontaneous polarization or magnetization components.

\begin{figure}[htb]
\includegraphics[width=\columnwidth]{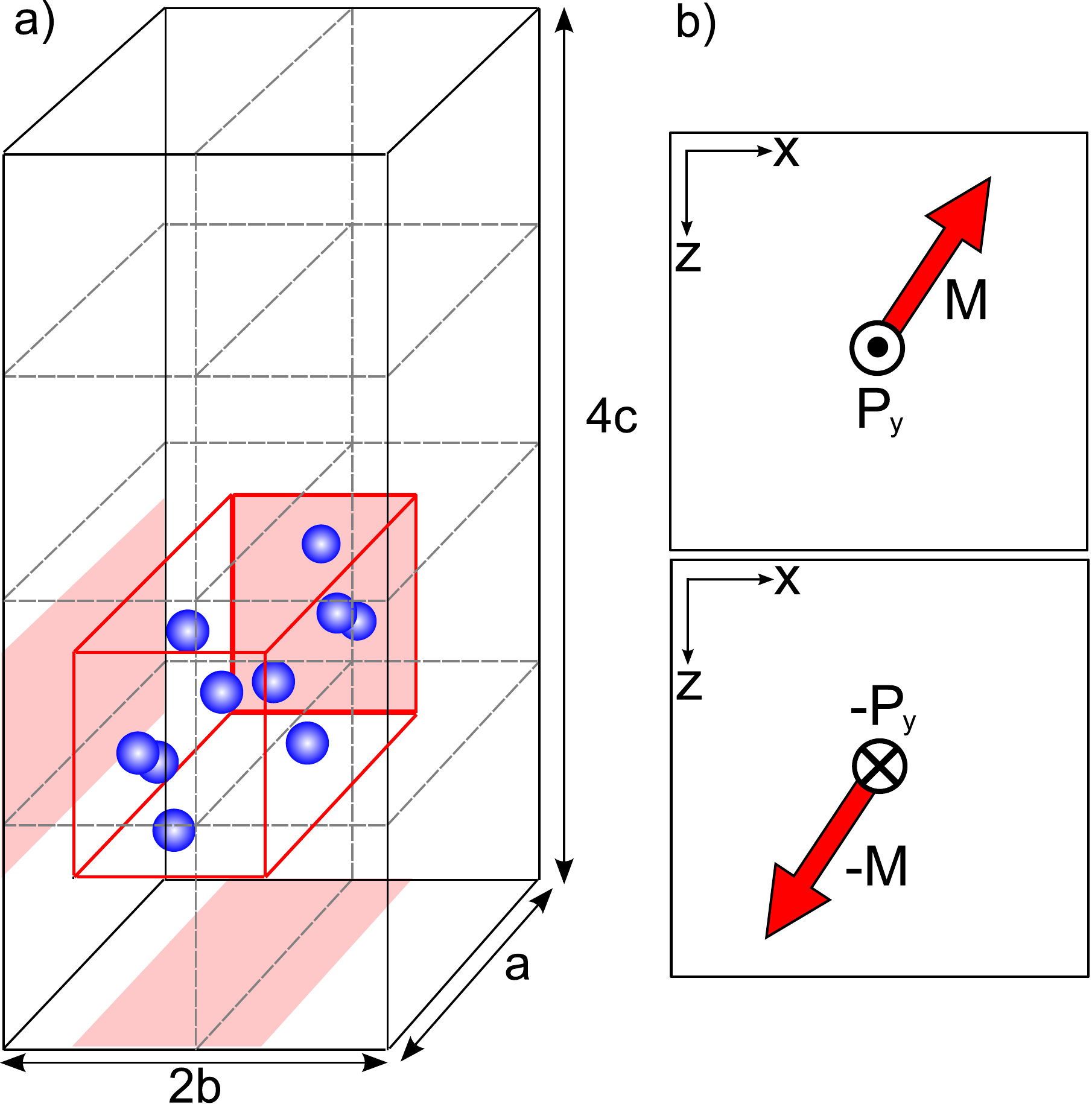}
\caption{\label{fig1} (a) Paramagnetic unit-cell embedded into the 16-fold unit-cell of the
multiferroic phase III of Co$_3$TeO$_6$. (b) Ferroelectric and weak ferromagnetic domains in phase
III. In the text we use Cartesian instead of the monoclinic coordinates according to $x\sim a$,
$y\sim b$, $z\sim c$.}
\end{figure}

\begin{figure}[htb]
\includegraphics[width=0.65\columnwidth]{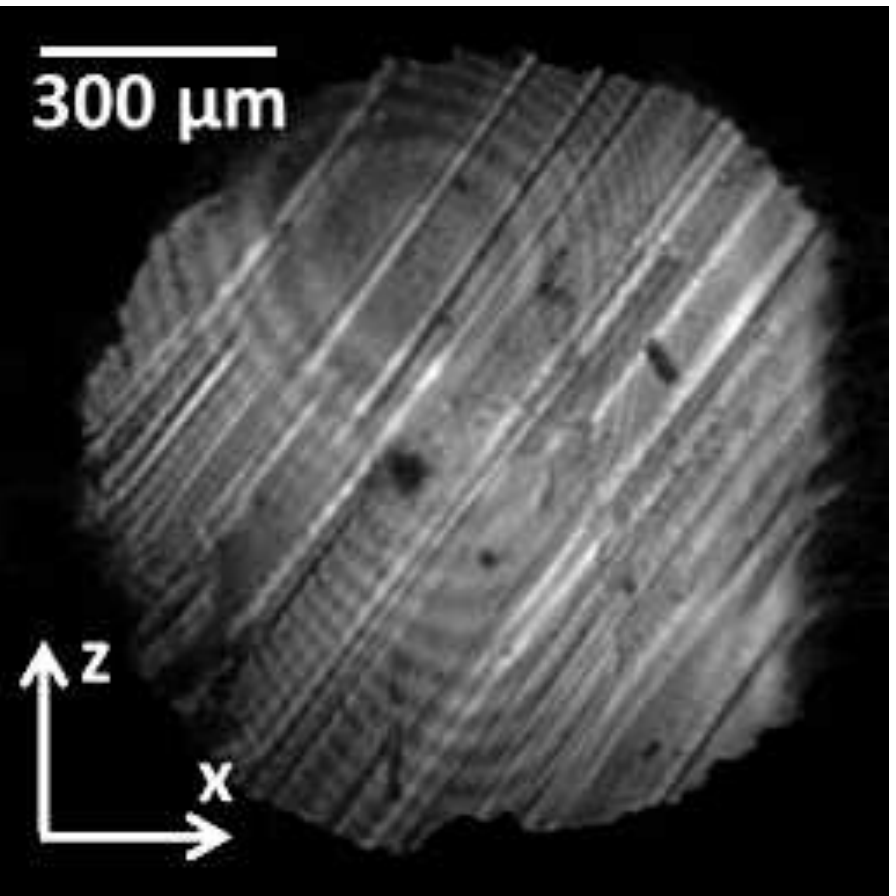}%
\caption{\label{fig2} SHG image of a polished single-crystal platelet (thickness $\unit[50]{\mu
m}$) of Co$_3$TeO$_6$ at $T = \unit[5]{K}$. Diagonal parallel stripes correspond to magnetic
domain walls, whereas the gradual ringlike oscillation of brightness is caused by interference of
the laser light in the sample. The incident light at $E=\unit[1.29]{eV}$ was propagating parallel
to the crystallographic $y$ axis and polarized parallel to the direction of the domain walls,
while the SHG light was polarized perpendicular to the domain walls.}
\end{figure}

III. At $T_2 = \unit[17.4]{K}$ a commensurate phase III arises in which the neutron diffraction
pattern corresponding to $\vec{k}_2$ persists coexisting with magnetic peaks associated with the
commensurate wave-vector $\vec{k}_3$. Both sets of reflections are observed in the whole range of
stability of the phase down to $\unit[1.6]{K}$.\cite{6} $\vec{k}_3$ is in
general a position inside the Brillouin-zone, corresponding to a four-dimensional IR of $G_P$,
denoted $\tau_1(\vec{k}_3)$, whose matrices are listed in Table~\ref{table1}. Keeping for the four
order-parameter components the same notation $(\eta_i = \rho_i e^{\pm i \theta_i}, i = 1,2)$ as
for phase I the transition free-energy reads
\begin{eqnarray}
 F_3(\rho_1, \rho_2, \theta_1, \theta_2) = F_1(\rho_1, \rho_2) + \nonumber\\
 \frac{\beta_3}{4}(\rho_1^4 \cos{4 \theta_1} + \rho_2^4 \cos{4 \theta_2}) + \\
 \frac{\beta_4}{2} \rho_1^2 \rho_2^2 \sin{2 (\theta_1 + \theta_2)} + \dots \nonumber\quad,
\end{eqnarray}
which differs from $F_1$ by the $\beta_3$ and $\beta_4$ lock-in invariants. Phase III results from
the coupling of the order-parameters $\eta_i (\vec{k}_3)$ and $\zeta(\vec{k}_2)$ corresponding to
the total free-energy
\begin{equation}\label{eq4}
 F_T = F_3(\rho_i, \theta_i) + F_2(\zeta) - \frac{\delta}{2} \zeta^2 (\rho_1^2+\rho_2^2)\quad,
\end{equation}
where the $\delta$-term represents the lowest-degree coupling between the two order parameters.
Table~\ref{table2} lists the symmetries and equilibrium conditions of the seven possibly stable
magnetic phases resulting from the minimization of $F_T$. The phase with magnetic symmetry
$\text{C}2'$ and a sixteen-fold multiplication of the primitive paramagnetic unit cell, shown in
Fig.~\ref{fig1}(a), coincides unambiguously with the reported neutron diffraction
observations\cite{6} in phase III. It corresponds to the equilibrium values of the
order-parameters $\zeta \neq 0, \rho_1 = \rho_2 = \rho_e, \theta_1 = \theta_2 = \theta_e$ which
allow emergence of a spontaneous polarization $P_y$ and a spontaneous weak magnetization $\vec{M}
= (M_x, M_z)$, with, respectively, two ferroelectric and two weakly ferromagnetic domains
(Fig.~\ref{fig1}(b)).
\begin{table*}
\caption{\label{table2} Seven possible choices (a) of the magnetic space groups (b) derived from
the minimization of $F_T(\zeta,\rho_i,\theta_i)$ in Eq.~(\ref{eq4}). (c) Equilibrium values of the
order parameters. (d) Basic translations of the conventional monoclinic or triclinic unit cells.
(e) Multiplicity of the volume of the primitive paramagnetic unit cell. (f) Origin of the
coordinates.}
\begin{ruledtabular}
\begin{tabular}{lllllc}
(a) & (b) & (c) & (d) & (e) & (f)\\
1 & $\text{P}\bar{1}'$ & $\zeta \neq 0, \rho_1 \neq 0, \rho_2 = 0, \theta_1 = 0$ & & 4 & $(0,1/2,0)$\\
2 & $\text{P}\bar{1}'$ & $\zeta \neq 0, \rho_1 \neq 0, \rho_2 = 0, \theta_1 = \pi/4$ & $\left\{\begin{array}{l}(-1,0,0)\\(1/2,3/2,1)\\(1/2,1/2,-1)\end{array}\right.$ & 4 & $(1/4,3/4,0)$\\
3 & $\text{P}1$ & $\zeta \neq 0, \rho_1 \neq 0, \rho_2 = 0$ & & 4 & $(0,0,0)$\\
4 & $\text{C}2'$ & $\zeta \neq 0, \rho_1 = \rho_2, \theta_1 = \theta_2$ & $(0,0,4),(0,2,0),(-1,0,0)$ & 16 & $(0,5/8,5/4)$\\
5 & $\text{P}\bar{1}'$ & $\zeta \neq 0, \rho_1 \neq 0, \rho_2 \neq 0, \theta_1 = 0, \theta_2 = -\pi/4$ & & 8 & $(0,1/2,0)$\\
6 & $\text{P}\bar{1}'$ & $\zeta \neq 0, \rho_1 \neq 0, \rho_2 \neq 0, \theta_1 = \pi/2, \theta_2 = -\pi/4$ & $\left\{\begin{array}{l}(0,2,0)\\(1,0,0)\\(0,1,-2)\end{array}\right.$ & 8 & $(1/4,1/4,-1/2)$\\
7 & $\text{P}1$ & $\zeta \neq 0, \rho_1 \neq 0, \rho_2 \neq 0$ & & 8 & $(0,0,0)$
\end{tabular}
\end{ruledtabular}
\end{table*}
Figure~\ref{fig2} shows the distribution of magnetic domains in the monoclinic $xz$ plane. The
image was gained by optical second harmonic generation (SHG) as described in Ref.~\onlinecite{5}.
The orientation of the domain walls along an arbitrary direction in the $xz$ plane further
confirms the $2^{\prime}$ symmetry. This symmetry is also consistent with the presence of the
$\chi_{xxx}$ and $\chi_{zzz}$ components of the SHG susceptibility tensor.\cite{5} Note, however,
that it differs from the symmetry $m$ assumed in Ref.~\onlinecite{5}. In Ref.~\onlinecite{5} only
i-tensor components were considered as origin of the SHG signal since in magnetically induced
ferroelectrics like MnWO$_4$ the SHG signal is always linearly coupling to the spontaneous
polarization. In contrast, the SHG signal leading to Fig.~\ref{fig2} revealed that SHG in CTO is
related to c-tensor components reproducing the weakly ferromagnetic order. SHG with
$\chi_{xxx}\neq 0$, $\chi_{yyy}=0$, and $\chi_{zzz}\neq 0$ as c-type susceptibilities leads to the
magnetic symmetry $2^{\prime}$. Unfortunately, a recent discussion of our SHG data in
Ref.~\onlinecite{16} is still based on the assumption of SHG coupling to the electric polarization
which lead to results inconsistent with the ones reported in this work.

Since the transitions to phases I and III result from the coupling of two order-parameters
corresponding to distinct $k$-vectors they display necessarily a first-order character\cite{10}
following the triggering mechanism proposed by Holakovsky\cite{11} in which one order-parameter
triggers the onset of another order-parameter across the first-order discontinuity. The mechanism
requires taking into account a sixth-degree invariant of the ``triggered'' order-parameter
$(\frac{\gamma}{6}(\rho_1^2+\rho_2^2)^3)$ in the total free-energy $F_{T}$. Under the conditions
$\delta > \left(\frac{|\beta_1+\beta_2|\lambda_1}{2}\right)^{1/2}$ and
$\lambda_1-\frac{4\delta^2}{|\beta_1+\beta_2|}<-4\left(\frac{\alpha_2\gamma}{3}\right)^{1/2}$ ,
the phase corresponding to the sole ``triggering'' order-parameter $\zeta$ becomes unstable with
respect to a phase in which both order parameters, $\zeta$ \textit{and} $\eta\sim
(\rho_i,\theta_i)$, are non-zero. In this phase $\zeta$ is frozen and $\eta$ determines the
symmetry breaking process. The transition can be shown to occur discontinuously at a higher
temperature than the transition temperature at which a phase with $\zeta=0, \rho_i\not=0$ would
appear.\cite{11} Accordingly, the triggering process, which is activated in the region of phase
coexistence preceding the transition at $T_2$, is due to a large \textit{negative} value of the
interaction term for the coupling between the two order-parameters, which determines the value of
the coupling coefficient $\delta$. Note that the first-order character of the transition to the
multiferroic phase at $T_2$ is confirmed by the strong discontinuities (of the order of
$\unit[10^{-2}]{\text{\AA}}$) observed in the lattice parameters\cite{6} and by a remarkably sharp
peak of the specific heat.\cite{5,6,12} In contrast, the transition from the paramagnetic to the
antiferromagnetic phase I is weakly first-order with almost negligible lattice discontinuities and
a standard specific heat anomaly.\cite{5,6} The transition from phase I to phase II which involves
a single order-parameter has typical second-order transition features with no noticeable
discontinuity of the lattice parameters.\cite{5,6}

The dielectric contribution to the free-energy $F_D = -\nu P_y \zeta^2 (\rho_1^2 \rho_2^2 \cos{2
(\theta_1+\theta_2)}) + \frac{P_y^2}{2 \epsilon_{yy}^0}$ yields the equilibrium value of $P_y$
below $T_2$:
\begin{equation}
 P_y = \nu \epsilon_{yy}^0 \zeta^2 \rho_e^4 \cos{4 \theta_e}
\end{equation}
At $T = T_2$, $P_y$ undergoes an upward discontinuity, imposed by the first-order character of the
transition. On further cooling it increases as $\approx (T_2-T)^2$, since the $\zeta$
order-parameter is frozen in phase III. A similar temperature dependence holds for the spontaneous
magnetization components $M_x$ and $M_z$. From the spontaneous magnetic contribution to the
free-energy in phase III $F_M = \mu_u M_u \zeta \rho_1^2 \rho_2^2 \cos{2 (\theta_1+\theta_2)} +
\frac{M_u^2}{2 \chi_{uu}^0}$ (with $u = x,z$), one gets
\begin{equation}
 M_u=-\chi_{uu}^0 \mu_u \zeta \rho_e^4 \cos{4 \theta_e}
\end{equation}
as the equilibrium value below $T_2$. Application of magnetic fields along $x$ or $z$ yields a
renormalization of the transition temperature according to
\begin{equation} \label{eq6}
 T_2(H_u) = T_2(0) - \chi_{uu}^0 \mu_u \alpha_0^{-1}\zeta^2 H_u^2\quad.
\end{equation}
Where $\alpha = \alpha_0 (T-T_2(0))$ is the coefficient of $\rho_e^2$ in $F_T$. For $\mu_u>0$ the
transition temperature is lowered under application of a magnetic field and $T_2(0)-T_2(H_u)$
increases quadratically with $H_u$, as observed in Co$_3$TeO$_6$ under $H_z$ field.
The magnetic susceptibility components $\chi_{uu}=\frac{M_u}{H_u}$ are obtained by minimizing the
field-induced contribution to the free-energy $F_M(H_u)=\mu_u M_u H_u \zeta^2 \rho_e^2 +
\frac{M_u^2}{2\chi_{uu}^0} - H_u M_u$. This reveals
\begin{equation}
 \chi_{uu}=\chi_{uu}^0 (1 - \mu_u \zeta^2 \rho_e^2)\quad.
\end{equation}

Therefore the discontinuous \textit{jump} of the order-parameter $\rho_e(T_2)$ coincides with a
\textit{drop} of $\chi_{uu}(T_2)$ the magnitude of which decreases with increasing field up to a
threshold field corresponding to $\rho_e(H_u^c)=\frac{1}{\zeta\sqrt{\mu_u}}$ above which
$\chi_{uu}$ undergoes an \textit{upward} discontinuity at $T_2(H_u)$. This behaviour is verified
experimentally\cite{5} for $H_z$ with a threshold field $H_z^c \approx \unit[12]{T}$. The slight
decrease with temperature observed for $\chi_{zz}$ \textit{below} $T_2$ indicates a slight
increase of the order-parameter within the multiferroic phase, with an almost step-like dependence
on temperature across the transition. As a consequence of this Heaviside-like behaviour the
specific heat $C$ which is proportional to the order-parameter derivative displays a delta-shape
like behaviour across $T_2$.\cite{5,6,12} This remarkable property of the first-order multiferroic
transition in Co$_3$TeO$_6$, which has been previously observed at the first-order ferromagnetic
transition in Fe$_2$P,\cite{13, 14} is reminiscent of structural transitions having a
reconstructive mechanism\cite{10} as observed, for example, at the fcc-hcp transition in
cobalt.\cite{15} However, in Co$_3$TeO$_6$ the symmetry-breaking mechanism involved at $T_2$ is
not of the reconstructive type since the $\text{C}2'$ symmetry of phase III is group-subgroup
related to the $\text{C}2'/\text{c}$ symmetry of phase II. Therefore the strong discontinuities
reported for the lattice parameters\cite{6} at $T_2$ should correspond to a strong magnetoelastic
coupling with an exchange energy critically sensitive to the interatomic spacing. This is in
agreement with the large spin lattice coupling deduced by Her et al.\cite{12} from the magnetic
hysteresis curves and with the direct exchange pathway corresponding to the shorter Co-Co
distances existing in phase III (see Fig.~12 in Ref.~\onlinecite{6}) for the neighbouring atoms
Co(5)-Co(5), Co(2)-Co(5), Co(3)-Co(3) and Co(4)-Co(4). The strong magnetoelastic effect is favored
by the coupling between the order-parameters $\zeta$ and $\eta_i$. Here, the already existing
antiferromagnetic order-parameter $\zeta$ triggers the emergence of the order parameter $\eta_i$
which is reflected by the transition discontinuity.

The respective monoclinic symmetries of phases II and III permit a variety of magnetoelectric
effects under applied magnetic or electric fields. For instance, in phase II applying a magnetic
field $H_y$ induces contributions by $P_x$ and $P_z$ to the polarization which vary as $P_{x,z}
\approx \zeta H_y$. In phase III one has $P_{x,z} \approx \zeta \rho_e^2 H_y$, i.e., at constant
temperature $P_x$ and $P_z$ increase linearly with $H_y$ in both phases, while at constant field
they increase with temperature as $(T_1-T)^{1/2}$ in phase I and as $(T_2-T)$ in phase III.
Conversely, applying the magnetic field $H_x$ or $H_z$ in phase II induces a polarization
polarization $P_y \approx \zeta H_{x,z}$. In phase III an additional contribution to the
spontaneous polarization $P_y$, namely $\Delta P_y \approx \zeta \rho_e^2 H_{x,z}$ is generated.
Reversed magnetoelectric effects should also be observed in phases II and III with the onset of an
induced magnetization $M_y$ depending linearly on electric fields $E_x$ or $E_z$. Additional
contributions to the $M_x$ and $M_z$ components of the spontaneous magnetization are generated
under an electric field $E_y$. Accordingly, our theoretical analysis suggests that
magnetic-field-induced polarization components $P_x(H_x)$, $P_x(H_z)$ and $P_z(H_x)$ are not
allowed. Their observation was reported earlier \cite{5} and should be related to a misorientation
of the sample admixing a $y$ component to the $x$ axis. The possibility for such a misorientation
was indeed pointed out by the authors of Ref.~\onlinecite{5}.

In summary, the observed properties of the sequence of three phases reported in Co$_3$TeO$_6$ have
been described theoretically. The most striking feature of this phase sequence is the coexistence
of propagation vectors in the incommensurate and commensurate multiferroic phases, the Brillouin
zone-centre $\vec{k}_2$ vector, which persists in all three phases, coupling successively with the
incommensurate $\vec{k}_1$-vector in phase I, and with its lock-in commensurate variant
$\vec{k}_3$ in phase III. Another remarkable property is the strongly first-order-character of the
multiferroic transition which relates both to the triggering mechanism coupling the $\vec{k}_2$
and $\vec{k}_3$ related order-parameters, and to a strong magnetoelastic coupling. The sharp peak
of the specific heat is shown to be consistent with the almost constant value of the spontaneous
magnetization in the multiferroic state. Furthermore, a number of experimental observations of
Ref.~\onlinecite{5} have been scrutinized: The monoclinic symmetry of phase III is $2^{\prime}$
(instead of the $m$ as previously proposed\cite{5}). This permits a spontaneous weak magnetization
($M_x$, $M_z$), which is found to be in agreement with domain structures observed in SHG
measurements. It also allows a spontaneous polarization $P_y$ which has not been investigated
before whereas contributions $P_{x,z}\neq 0$ are no longer expected. The magnetoelectric effects
that should exist in Co$_3$TeO$_6$ have been worked out theoretically. A verification of these
predictions, supported by the application of both magnetic and electric fields, is necessary for
confirming the validity of the theoretical description presented in this article.

At last we emphasize that our theoretical analysis and conclusions differ in an essential manner
from the symmetry analysis proposed in Ref.~\onlinecite{16} for two reasons. (i) Scrutinization of
the data reported in Ref.~\onlinecite{5} lead to a revision of the magnetic symmetry and the
direction of electric polarization. Since this is a very recent result it could be taken into
account in the present work but not in Ref.~\onlinecite{16}. (ii) In Ref.~\onlinecite{16} mainly
one dimensional order-parameters associated with the wave-vector $\vec{k}_2$ are considered. In
contrast, we analyze the symmetries and physical properties of phases I and III as resulting from
the coupling of order-parameters associated respectively with the wave-vectors ($\vec{k}_1$,
$\vec{k}_2$) and ($\vec{k}_2$, $\vec{k}_3$), in agreement with the neutron diffraction data
reported in Ref.~\onlinecite{6}. In particular the field-induced component $P_z(H_x)$ discussed in
Ref.~\onlinecite{16} is shown to be absent in our description, whereas we predict the existence of
a single spontaneous (zero-field) polarization component $P_y$.

\begin{acknowledgments}
The authors thank Roland Mathieu and Per Nordblad for helpful discussions, and the G\"{o}ran
Gustafsson foundation and the SFB 608 of the DFG for financial support.
\end{acknowledgments}

\bibliography{toledano}

\begin{thebibliography}{17}
\bibitem [{foo()}]{footnote-CTO}%
  \BibitemOpen
  \href@noop {} {}\bibinfo {note} {Shifted and additional transition
  temperatures are reported in magnetic-field-dependent measurements in
  Ref.~\onlinecite{6}. However, just like Ref.~\onlinecite{16} we base our
  theoretical analysis on the data reported in
  Refs.~\onlinecite{5,6}.}\BibitemShut {Stop}%
\end{thebibliography}

\end{document}